\documentclass[12pt,amsmath,amssymb,unsortedaddress,superscriptaddress,prb]{revtex4-1}
\usepackage{subeqn}
\usepackage{graphicx}
\usepackage{dcolumn}
\usepackage{bm}
\usepackage{pifont}

\begin{document}

\title{Intrinsic and dopant enhanced solid phase epitaxy in amorphous germanium}
\author{B. C. Johnson\footnote{Email: johnsonb@unimelb.edu.au}}
\affiliation{School of Physics, University of Melbourne, Victoria 3010, Australia}
\author{P. Gortmaker}
\affiliation{Department of Electronic Materials Engineering, The Australian National University, Canberra 0200, A.C.T., Australia}
\author{J. C. McCallum}%
\affiliation{School of Physics, University of Melbourne, Victoria 3010, Australia}

\begin{abstract}
The kinetics of intrinsic and dopant-enhanced solid phase epitaxy (SPE) is studied in amorphous germanium ($a$-Ge) layers formed by ion implantation on $<$100$>$ Ge substrates. The SPE rates were measured with a time-resolved reflectivity (TRR) system between 300 and 540~$^{\circ}$C and found to have an activation energy of (2.15$\pm$0.04)~eV. To interpret the TRR measurements the refractive indices of the $a$-Ge layers were measured at the two wavelengths used, 1.152 and 1.532~$\mu$m. For the first time, SPE rate measurements on thick $a$-Ge layers ($>$3{~$\mu$m}) have also been performed to distinguish between bulk and near-surface SPE growth rate behavior. Possible effects of explosive crystallization on thick $a$-Ge layers are considered. When H is present in $a$-Ge it is found to have a considerably greater retarding affect on the SPE rate than for similar concentrations in $a$-Si layers. Hydrogen is found to reduce the pre-exponential SPE velocity factor but not the activation energy of SPE. However, the extent of H indiffusion into a-Ge surface layers during SPE is about one order of magnitude less that that observed for $a$-Si layers. This is thought to be due to the lack of a stable surface oxide on $a$-Ge. Dopant enhanced kinetics were measured in $a$-Ge layers containing uniform concentration profiles of implanted As or Al spanning the concentration regime 1--10~$\times$10$^{19}$~/cm$^{-3}$. Dopant compensation effects are also observed in $a$-Ge layers containing equal concentrations of As and Al, where the SPE rate is similar to the intrinsic rate. Various SPE models are considered in light of these data.

\end{abstract}
\pacs{81.15.Np,81.15.Aa,61.72.uf}
\maketitle

\section{Introduction}
Crystallization of ion implanted materials via solid phase epitaxy (SPE) is a common processing step during device fabrication due to its ability to achieve high dopant activation with a low thermal budget.\cite{itrs} A fairly extensive literature exists based on SPE studies with amorphous Si. However, only a few researchers have reported on SPE measurements in Ge.\cite{SSC:csepregi,don,lu} Due to recent developments in nano-scale electronics and opto-electrical devices, Ge has regained some interest.\cite{depuydt:mssp} The need for current and accurate SPE data in amorphous Ge is now quite apparent. Furthermore, Ge is an ideal alternative to Si in which to gain further insight into the SPE process and the strengths and limitations of the various SPE models.\cite{lu}

SPE is a thermally activated process and the velocity of the crystalline--amorphous ({\em c--a}) interface through the amorphous phase can be described by an Arrenhius-type equation of the form,

\begin{equation}
v = v_{o} \; {\rm e}^{(-E_{a}/kT)}
\label{spe}
\end{equation}
 
\noindent where $v_{o}$ and $E_{a}$ are the velocity pre-exponential factor and activation energy of SPE, respectively. For 0.5~$\mu$m thick amorphous Ge ($a$-Ge) layers, Csepregi~{\em et~al.\/} reported SPE rates over the temperature range 310--340~$^{\circ}$C using partial furnace annealing combined with Rutherford backscattering spectroscopy and ion channeling (RBSC) measurements.\cite{SSC:csepregi} They reported an activation energy of 2.0~eV. Donovan~{\em et~al.} used calorimetry measurements to measure the heat of crystallization in the temperature range 417--457~$^{\circ}$C.\cite{don} From a fit to the normalized power output of the calorimeter as a function of temperature they reported an activation energy of 2.17~eV and a velocity prefactor of $4 \times 10^{7}$~m/s. Lu~{\em et~al.} have measured the SPE rate of 0.8~$\mu$m thick $a$-Ge layers over the temperature range 300--365~$^{\circ}$C using a time resolved reflectivity (TRR) apparatus similar to the system used in the present work.\cite{Lu:1990ab,lu} They reported an activation energy of 2.17~eV and a velocity prefactor of $1.2\times 10^{7}$~m/s.

Interest in Si$_{x}$Ge$_{1-x}$ alloys has prompted researchers to measure SPE rates for $a$-Ge, $a$-Ge being the $x=0$ end point of the alloy curve.\cite{krin,hay} For example, Olson and Roth have reported an activation energy of 2.26~eV in a temperature range 350--450~$^{\circ}$C for $a$-Ge layers formed by Si implantation.\cite{review:roth} Using amorphous layers 0.09--0.28~$\mu$m thick, Kringh$\o$j~{\em et~al.} have reported an SPE activation energy of 2.02~eV with a velocity prefactor of $6.1\times 10^{6}$~m/s.\cite{krin} Haynes~{\em et~al.}, using SPE rate data over the depth range 0.08--0.16~$\mu$m and over the temperature range 290--390~$^{\circ}$C, reported a value of 2.19~eV and $7\times 10^{7}$~m/s for the activation energy and velocity prefactor, respectively.\cite{hay} 

The activation energies in these studies range between 2.0 and~2.3~eV with a velocity prefactor lying between $6.1\times 10^{6}$ and $7\times 10^{7}$~m/s. These Arrhenius factors for the SPE rate in $a$-Ge are not yet known to an accuracy comparable to the corresponding Si values which are generally accepted to be 2.7~eV and $4.64\times 10^{7}$~m/s, respectively.\cite{APL:roth} Indeed, Lu~{\it et~al.\/} noted that the uncertainty in their Ge SPE data results in an uncertainty of a factor of $\sim$50 for the crystallization factor calculated from their extended kinetic model of the growth process.\cite{lu} Furthermore, the thickest $a$-Ge layers used to date in SPE measurements were 0.8~$\mu$m thick, while the majority of the measurements involved layers 0.5~$\mu$m or less in thickness. Roth~{\em et~al.\/} have demonstrated that hydrogen contamination can effect the SPE rate in Si for interface depths up to 2~$\mu$m.\cite{APL:roth,MRSSP:roth92} Atomic H is formed as a by-product of the oxidation at the $a$-Si surface during annealing in the presence of water vapor. Once H is inside the $a$-Si layer it diffuses rapidly and interacts with the {\em c--a} interface. Furthermore, even when anneals are performed in vacuum, H present in the surface oxide diffuses into the amorphous layer.\cite{MRSSP:roth92} Before the current study it was unclear whether the existing thin layer Ge SPE measurements were afflicted by the same problem.

The very mechanism by which atoms rearrange during SPE is still an area of considerable debate.  Lu {\em et al.} have established that the SPE rates in Si and Ge are enhanced by pressure and are characterized by negative activation volumes of $\Delta V^{*}_{Ge} = -0.45 \Omega_{Ge}$ and $\Delta V^{*}_{Si} = -0.28 \Omega_{Si}$, where $\Omega_{Ge}$ and $\Omega_{Si}$ are the atomic volumes of crystal Ge and crystal Si, respectively.\cite{lu} This data together with the positive activation volume for Ge self-diffusion is cited as evidence that the transport of vacancies to the {\em c--a} interface is not the rate-limiting step in Ge SPE. 

Other studies in $a$-Si have shown that the SPE rate is sensitive to shifts in the Fermi level caused by the presence of dopants and that both neutral and charged defects may be responsible for the SPE process.\cite{will:role,lu,APL:mccallum,TOBE:gfls} There have been very few studies on the SPE rates in dopant implanted $a$-Ge. However, such information could be an important key to understanding the atomic rearrangement processes responsible for SPE. Suni~{\em et~al.\/} have observed enhanced SPE in B and As implanted $a$-Ge.\cite{suni2} Using the furnace/RBSC technique the SPE rates were observed to be 1.5 and 2.5 times faster than the intrinsic rate, respectively. The SPE rate was found to return to its intrinsic rate when similar concentrations of p-type and n-type dopants were present. However, this work is not quantitative enough to allow any substantial conclusions to be drawn about dopant-enhanced SPE in $a$-Ge. 

There are a number of molecular dynamics (MD) studies which attempt to simulate the motion of the {\em c--a} interface in Si during SPE on an atomic scale.\cite{saito81,saito84,prb:bernstein98,prb:bernstein00,PhysRevB.61.8537,prb:mattoni} Some of these studies show fair agreement with the experimentally determined activation energy of SPE in Si.\cite{PhysRevB.61.8537,krzeminski:123506} Several possible SPE mechanisms have also been identified. To our knowledge no SPE MD simulations for Ge exist at present. 

In this paper, we present comprehensive SPE measurements for intrinsic $a$-Ge formed on $<$100$>$ Ge substrates by Ge ion implantation. The growth kinetics have been measured over a temperature range of 300--540{$\;^{\circ}$C}, which is substantially greater than that used in other SPE measurements in $a$-Ge. These results are presented in section~\ref{intrinsic}.  For the first time, comparisons have been made between the SPE rates in thick $a$-Ge layers ($>$3~$\mu$m) and thinner layers ($\sim$1.5~$\mu$m thick) to distinguish between bulk and near-surface SPE growth rate effects. This is presented in section~\ref{thick}. These measurements have allowed us to identify and quantify the effects of H infiltration during SPE and measure the H-free intrinsic SPE rate for the first time allowing us to determine the most accurate activation energy and prefactor for the process. These results explain the scatter in the values obtained by previous authors where thin $a$-Ge layers were used exclusively. The effect of H on the SPE rate is studied in detail with H implanted $a$-Ge in section~\ref{hydrogen}.

We also present new data in section~\ref{dopant} showing that dopant-enhanced SPE in $a$-Ge occurs for concentrations of implanted As and Al greater than $1 \times 10^{19}$ cm$^{-3}$ and that it exhibits a similar dependence on concentration and temperature to that observed for dopant-enhanced SPE in H-free $a$-Si layers.\cite{APL:mccallum,TOBE:gfls} SPE growth models are considered in light of these data in sections~\ref{kinetic} and~\ref{gfls}. The generalized Fermi level shifting model shows excellent agreement with previous results obtained in Si and with an SPE mechanism based on the dangling bond type defect. An Appendix is included which outlines the methods used to implement this model for $a$-Ge layers in the temperature and concentration regime used in this work. Finally, links between the present study and MD simulations as a possible means of identifying the mechanism giving rise to dopant enhancement is discussed in detail.

\section{Experimental}

\subsection{Sample Preparation}

The kinetics of intrinsic and dopant enhanced SPE were measured in $a$-Ge layers formed by self-ion implantation into Ge $<$100$>$ wafers. Wafers from different suppliers, and with different background doping levels were used in an effort to determine if the origin of the material had any effect on the SPE rate.  The Ge wafers included \mbox{p-type}~(1--5~$\Omega\cdot$cm), \mbox{n-type}~($<$0.4~$\Omega\cdot$cm), undoped~($>$30~$\Omega\cdot$cm), and undoped~($>$20~$\Omega\cdot$cm, 0.5~mm wafers) material.  While true intrinsic germanium has a resistivity of 47~$\Omega\cdot$cm,\cite{sze} the background doping levels in these wafers were not expected to have a measurable effect on the SPE rate. Indeed, dopant-enhanced SPE is not observed until the dopant concentration is greater than $\sim$1~$\times$10$^{19}$~/cm$^{3}$ as reported in the present work. 

A National Electrostatics Corp.\ 1.7~MV tandem accelerator was used for all implants. During implantation, substrates were affixed with Ag paste to the implanter stage, which was held at 77~K. The samples were tilted 7$^{\circ}$ off the incident beam axis to avoid channeling and also rotated about the surface normal by a similar amount to prevent any remaining possibility of planar channeling.\cite{armi}

Sequential implants at 0.55, 1.0 and 2.0 MeV, each to a fluence of $\rm 5\times10^{15}\; Ge/cm^{2}$, were used to create $a$-Ge layers $\sim$1.5~$\mu$m thick. One set of samples was created with the same sequence of energies but with only 20\% of the fluence to investigate any possible dependence of the SPE rate on the amorphization fluence. Multiple energy implants at 0.8, 2.0, 4.6 and 7.6~MeV, each to a fluence of 1.5{$\times$10$^{15}\:$Ge/cm$^{2}$}, were also used to create $a$-Ge layers $\sim$3.25~$\mu$m thick. Neither visual inspection nor RBS measurements showed any evidence of the porous $a$-Ge structure that has been reported for high-fluence room temperature implanted Ge.\cite{appl2,holl}

Secondary ion mass spectroscopy (SIMS) was performed on selected samples to measure the H concentration profile in the near surface region after a partial anneal. SPE rates and H profiles were also measured in $a$-Ge layers implanted with 80~keV H to fluences of $\rm 3\times 10^{14}\; /cm^{2}$, $\rm 6\times 10^{14}\; /cm^{2}$ or $\rm 1.5\times 10^{15}\; /cm^{2}$, that formed Gaussian-like concentration profiles centered at 0.68~$\mu$m.

For the dopant-enhanced SPE studies, multiple energy implants at 77~K into pre-amorphised samples were used to produce uniform As and Al concentration profiles over the depth range 0.25--0.55~$\mu$m and 0.5--0.9~$\mu$m, respectively. Fluences were chosen to result in peak concentrations of 1, 5 and 10~$\times$10$^{20}$~/cm$^{3}$. An additional sample was implanted with both As and Al to concentrations of 5~$\times$10$^{19}$~/cm$^{3}$ for dopant compensation measurements. The depth profiles of As and Al were also measured by SIMS.

\subsection{Time Resolved Reflectivity}

The SPE rates of the {\em c--a} interface were determined from time resolved reflectivity (TRR) measurements by acquiring reflectivity data simultaneously using  two HeNe lasers at wavelengths of $\lambda = 1.152$~$\mu$m to probe the 1.5~$\mu$m thick $a$-Ge layers and at $\lambda = 1.523$~$\mu$m for the thicker ($\sim$3.25~$\mu$m) layers. As the {\em c--a} interface moves through the sample, peaks in the TRR reflectivity trace occur every $\lambda /2n $. By combining the measured TRR traces and a theoretical reflectivity versus amorphous thickness curve the velocity of the interface was determined. These data were collected while the samples were held on a resistively heated vacuum chuck and annealed in air over the temperature range 300--540~$^{\circ}$C. The temperature of the samples during the anneals was calibrated by comparing the reading of a type-K thermocouple embedded in the sample stage with the melting points of various suitably encapsulated metal films evaporated onto Si wafers. The error associated with the temperature reading was found to be $\pm 1\;^{\circ}$C. Measurements were performed in air to match the experimental conditions of other studies and so that the effects of H infiltration could be examined and quantified. Further details on the experimental apparatus are presented elsewhere.\cite{APL:mccallum} 

\section{Results and Discussion}
\subsection{Refractive index of $a$-Ge}
\label{refractive}

TRR experiments involving $a$-Ge usually rely on refractive index values for sputter deposited or evaporated films.\cite{con} Compared to $a$-Ge layers produced by ion implantation, the sputter deposited films typically suffer from density variations and relatively high levels of incorporated impurities such as oxygen. To determine an $a$-Ge refractive index value suitable for the ion implanted layers used in this study, samples were partially annealed for various times on the TRR system and then the thickness of the remaining $a$-Ge layer was measured with Rutherford backscattering spectroscopy and ion channeling (RBSC).\cite{paul} The $a$-Ge layer thickness was calculated by assuming that the density of $a$-Ge is the same as $c$-Ge. This is not unreasonable as the density of well-relaxed evaporated $a$-Ge films has been verified to be close to that of $c$-Ge.\cite{polk} However, for $a$-Si the density is typically 1.2\% less than that of the $c$-Si value.\cite{cus} If a similar reduction in density is observed in $a$-Ge formed by ion implantation then the calculated index of refraction would be over-estimated by the same percentage. However, this over-estimation is within the quoted error of the determined value. Also, the systematic use of an over-estimated index of refraction value will not affect the activation energy determined from the TRR data. However, it will affect the determined absolute value of the solid phase epitaxial regrowth rates.

The real part of the refractive index for $a$-Ge was determined to be $5.34\pm 0.15$ and $5.07\pm 0.17$ for the 1.152~$\mu$m and 1.523~$\mu$m lasers, respectively. The complex part of the latter was also measured to be $j(0.095\pm 0.008)$. Measurements in the temperature range 340--440~$^{\circ}$C indicated that the refractive index has no significant temperature dependence over the temperature range spanned by the SPE measurements. While the refractive index values reported here for $a$-Ge formed by ion implantation are somewhat higher than the values obtained by Connell~{\em et~al.}\ for sputter deposited and evaporated $a$-Ge films (e.g. $\sim$4.8 at 1.15{~$\mu$m}),\cite{con} this is consistent with the relatively high concentrations of impurities and voids expected to be present in their films.

\subsection{SPE in Intrinsic $a$-Ge}
\label{intrinsic}

\begin{figure}
\begin{center}
\rotatebox{0}{\includegraphics[height=12cm]{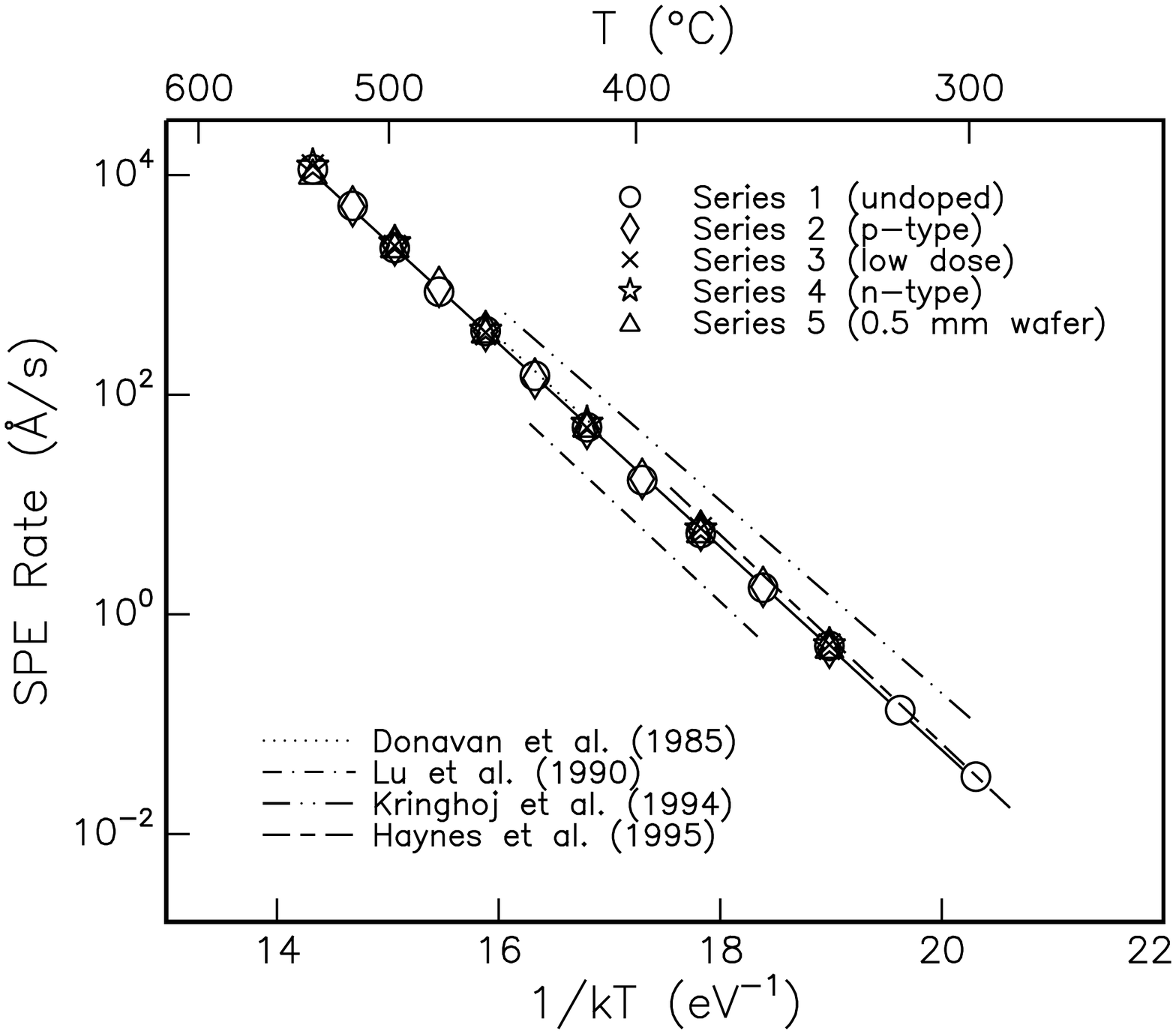}}
\end{center}
\caption[Arrhenius Behaviour for Various $a$-Ge  Substrates]{
The SPE rates for the various a-Ge layers displayed on an Arrhenius plot with fit giving E$_a$~=~2.15~eV and v$_o$~=~(2.6$\pm$0.5)$\times$10$^{7}$~m/s. Substrates are: Series~1, undoped~($\circ$), Series~2, p-type~($\Diamond$), Series~3, low-fluence amorphization~($\times$), Series~4, n-type~(\ding{73}), Series~5,~0.5~mm substrate~($\triangle$). Errors are $\pm 15 \%$ for the velocity values and  $\pm 1.5 \; ^{\circ}$C for the temperature values and are about the size of the symbols. The SPE rates determined from the activation energies and velocity prefactors reported by previous authors are also shown for comparison. These are plotted over the temperature range in which they were measured. 
}
\label{ge_arh2}
\end{figure}

The $\sim$1.5~$\mu$m thick surface $a$-Ge layers formed by Ge implantation were used to determine the SPE regrowth behavior of intrinsic high-purity $a$-Ge. TRR data were collected, using the 1.15~$\mu$m laser, in 20~$^{\circ}$C intervals from 300~$^{\circ}$C to~540~$^{\circ}$C. The SPE rate was extracted from the depth range over which it was constant. This was found to be the case for interface depths greater than 0.3~$\mu$m. The SPE rate versus temperature data are presented in Arrhenius form in Fig.~\ref{ge_arh2}. The errors were calculated by considering the reproducibility of the data and the RMS noise in the determined velocity curves and are about the size of the symbols. The average activation energy determined from these measurement sets is E$_a = (2.15\pm 0.04)$~eV and the velocity prefactor is v$_o$~=~(2.6$\pm$0.5)$\times$10$^{7}$~m/s. No statistically significant difference was observed in the SPE behavior of any of the intrinsic $a$-Ge sample sets containing different background doping, $a$-Ge layer thickness, or produced under different amorphisation conditions. The SPE rates determined from the activation energies and velocity prefactors reported by previous authors are also shown for comparison. These are plotted over the temperature range in which they were measured. Results reported by Csepregi~{\em et~al.} and Roth~{\em et~al.} are not shown since the velocity prefactor was not reported in their work.\cite{SSC:csepregi,review:roth} Given that our measurements were determined only where the SPE rate was constant, span a temperature range of 300--540~$^{\circ}$C which is 100~$^{\circ}$C greater than any of the other measurements and are based on an independent determination of the refractive index of $a$-Ge, we believe that the values reported here for the activation energy and prefactor for SPE of intrinsic $a$-Ge surface layers represent the most reliable data available. The variation in the values reported by other authors could be due to the limited temperature range spanned, use of an incorrect value of the refractive index, the formation of a-Ge layers using Si implants (creating an alloyed layer) or thin layer effects possible associated with H infiltration as observed in thin $a$-Si layers. The latter is discussed further in the next section.

\subsection{SPE in Thick $a$-Ge Layers}
\label{thick}

Amorphous Ge layers $\sim$3.25~$\mu$m thick were also studied to distinguish between bulk and near-surface SPE growth rate behavior. One difficulty in working with thick $a$-Ge layers is the increased probability that the layer will not crystallize by SPE but will instead undergo explosive or self-sustained crystallization.\cite{jack} In this process the annealing temperature combined with the heat release from crystallization are sufficient to maintain a melt-mediated growth process, once initiated. There is only enough energy available to sustain this phenomenon above some critical temperature, T$_{c}$, and it has been observed that T$_{c}$ generally decreases with increasing amorphous layer thickness.\cite{leam} The melt-mediated growth has been observed to proceed at rates in excess of 1~m/s.  Explosive crystallization was quite common in the $\sim$3.25~$\mu$m $a$-Ge layers and SPE measurements could only be performed for temperatures less than $\sim$440$\;^{\circ}$C. The source point for the explosive crystallization event was often a small chip in the cleaved edge of the sample. Those samples that cleaved cleanly had the highest probability of crystalizing by~SPE.

The SPE rates for the thick $a$-Ge layers were measured between 360--440$~^{\circ}$C.\cite{paul} While the temperature range is not as comprehensive as the study of the thinner $a$-Ge  layers described above, it is worth reporting the Arrhenius dependence for comparison. The activation energy determined from this data set was 2.16~eV, with a velocity prefactor of~3.3$\times$10$^{7}$~m/s. This value is in good agreement with the value previously determined from the~1.15~$\mu$m TRR study of thinner $a$-Ge layers.

The velocity of the {\em c--a} interface was found to remain essentially constant over the majority of the $\sim$3.25~$\mu$m depth range. No large scale velocity reductions were observed at or around the 2~$\mu$m mark, in contrast to the results reported by Roth~{\em et al.}\ for silicon.\cite{MRSSP:roth92} However, a slight velocity reduction during the final $\sim$0.4~$\mu$m of regrowth was visible, similar to that reported by Lu~{\em et al.}\ who speculated that it may be due to surface impurities known to retard SPE growth that have been driven into the sample during the multiple amorphization implants.\cite{lu} Olson and Roth also mention having observed a near surface reduction in the $a$-Ge SPE rate in some of their unpublished data, and they attribute this to hydrogen diffusing into the amorphous material.\cite{review:roth}

The lack of a large velocity change in the $a$-Ge system, as compared to the a-Si system could be due to a number of factors. For example, there may not be an intake of H from the environment into the $a$-Ge layer to large depths as observed in $a$-Si. Alternatively, H may not have a retardation effect on the SPE regrowth rate in $a$-Ge. Since the thick layer SPE data was not sufficient to determine which of these reasons correctly explains the lack of a significant H retardation of the SPE rate compared to that observed in Si, further measurements were undertaken. These included SIMS analysis of partially annealed samples, similar to the measurements performed by Roth{\em et al.},\cite{APL:roth} and also TRR and SIMS measurements on H implanted $a$-Ge layers. These results are presented in the next section.

\subsection{Hydrogen effects}
\label{hydrogen}

\begin{figure}
\rotatebox{0}{\includegraphics[height=12cm]{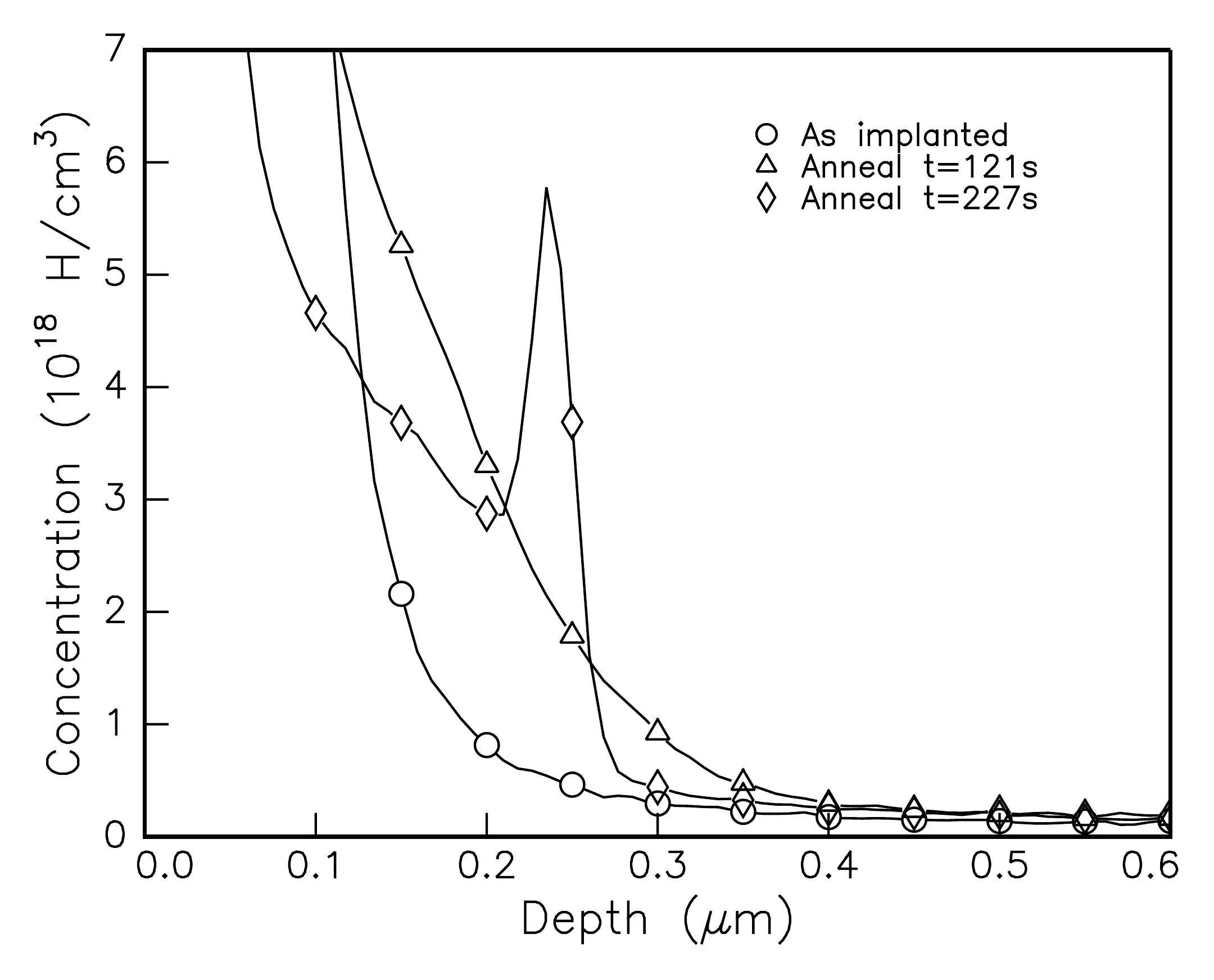}}
\caption[SIMS Profiles of~H Content in Partially Annealed $a$-Ge Samples]{
SIMS profiles of the hydrogen content in thick $a$-Ge samples: in the as-implanted state ($\circ$), and for partial anneals at 420~$^{\circ}$C for durations of 121~seconds ($\triangle$) and 227~seconds ($\diamondsuit$). The expected {\em c--a} interface depths for each of these partial anneals based on the TRR data were 0.79~$\mu$m and 0.24~$\mu$m, respectively.
}
\label{ge_h_sims1}
\end{figure}

Roth {\em et al.}\ observed an SPE rate retardation due to hydrogen infiltration during crystallization measurements on surface $a$-Si layers.\cite{APL:roth,MRSSP:roth92} This raises the question of whether a similar effect may occur in the SPE of surface $a$-Ge layers. To consider this issue,  SIMS was performed on the thick $a$-Ge layers described above at three different stages of partial annealing. Fig.~\ref{ge_h_sims1} shows the H concentration in these $a$-Ge layers. The first sample was analyzed in the as-implanted state to determine the background level of H and to verify that there was no hydrogen intake from the implantation process. The second sample was annealed for 121~seconds, and the third for 227~seconds both at 420~$\;^{\circ}$C. The expected {\em c--a} interface depths as determined from TRR data for these two partial anneals were 0.79$\pm$0.02~$\mu$m and 0.24$\pm$0.02~$\mu$m, respectively. 

The 121~second anneal was chosen so that if there was a surface-based source of H, its associated profile could be viewed prior to interaction with the {\em c--a} interface. It can be seen that the H concentration is at the measurement background level ($\sim$10$^{17}$~/cm$^{3}$) for most of the layer, but rises sharply from about 0.2~$\mu$m  through to the surface. At this point, the {\em c--a} interface has advanced 0.62~$\mu$m, still leaving 0.79~$\mu$m of $a$-Ge, so that the {\em c--a} interface is well beyond the range of the hydrogen. This is contrasted against the Si case, where Roth~{\em et~al.} reported H penetration to depths of 1~$\mu$m after only 0.2~$\mu$m of regrowth. After 1.4~$\mu$m of regrowth in the Si case, the H had penetrated to a depth of 2.7~$\mu$m.\cite{MRSSP:roth92}

For the 227 second anneal,  the {\em c--a} interface is at a depth of 0.24~$\mu$m and has come into contact with the indiffused H. The hydrogen content is observed to drop to background levels upon crossing from the $a$-Ge side of the interface to the $c$-Ge side. A peak in the~H concentration profile is formed on the $a$-Ge side of the interface as the hydrogen is progressively pushed ahead of the advancing {\em c--a} interface. Roth~{\em et~al.\/}\cite{MRSSP:roth92} observed similar zone-refinement of the H in Si by the {\em c--a} interface. The level of H as seen by the interface at this point in the anneal is~$\sim 6\times$10$^{18}$~/cm$^{3}$. This value is of the same order as reported by Roth, who observed values between 2$\times$10$^{18}$~/cm$^{3}$ and 1.5$\times$10$^{19}$~/cm$^{3}$, depending upon how far the interface was allowed to progress, and hence how much hydrogen it had collected.

These results show that hydrogen does diffuse into $a$-Ge  layers from the ambient during thermal processing, but that the depth range over which this effects SPE measurements is about one order of magnitude less than that which is observed in silicon. For $a$-Ge layers 1.5~$\mu$m thick, the SPE rate reduction associated with this indiffused H is observed to be limited to the first $\sim$0.2~$\mu$m. For thicker layers, the affected depth is expected to increase, as the H profile has a longer time to diffuse into the $a$-Ge layer before coming in contact with the {\em c--a} interface. This appears to hold true, as the velocity data for the 3.25~$\mu$m\ $a$-Ge layers were observed to exhibit an SPE rate reduction when the {\em c--a} interface reached to within $\sim$0.4~$\mu$m of the surface (data not shown).

Thus, all previous studies dealing with $a$-Ge layers are shown to be affected by H.\cite{SSC:csepregi,don,lu,Lu:1990ab,krin,hay,review:roth} In fact, two of these studies utilized data from entirely within 0.3~$\mu$m of the surface.\cite{krin,hay} Hence, it is unlikely that these works accurately represent the bulk SPE behaviour of $a$-Ge. In light of this, the SPE data used to determine the activation energy in the present work was taken from beyond this depth and hence encompassed the depth range over which the SPE rate was constant. 

\begin{figure}
\rotatebox{0}{\includegraphics[height=12cm]{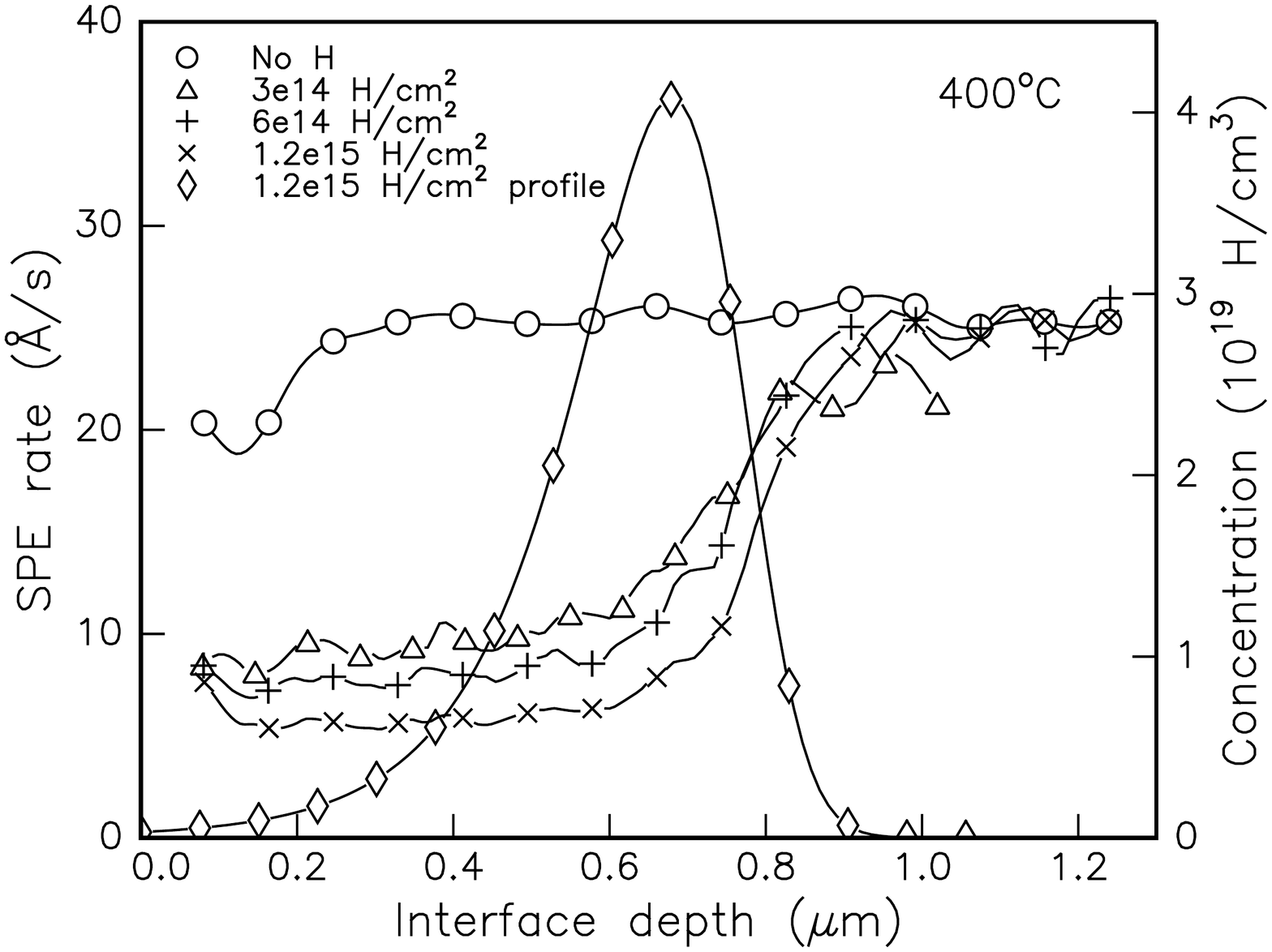}}
\caption[Effect of Implanted~H on a-Ge SPE Rate]{
The effect of~80~keV implanted hydrogen on the solid phase epitaxial 
regrowth rate in $a$-Ge, for hydrogen fluences of $\rm 3 \times 10^{14}\; /cm^{2}$,~$\rm 6 \times 10^{14}\; /cm^{2}$, and $\rm 1.2 \times 10^{15}\; /cm^{2}$. The theoretical as-implanted profile obtained from the Profile code\cite{profile} for the $\rm 1.2 \times 10^{15}\; /cm^{2}$  case is shown for comparison. The as-implanted profiles for the other two fluences would be identical except for a vertical scaling factor of~0.25 and~0.5, respectively.
}
\label{ge_h_vel}
\end{figure}

To determine whether H concentrations similar to those observed to retard SPE in $a$-Si layers have a similar effect in $a$-Ge, hydrogen implants were performed directly into the thin $a$-Ge. The implantation conditions were chosen to create peak hydrogen concentration levels comparable to those measured in silicon and reported by Roth~{\em et al}.\cite{MRSSP:roth92} Three implantation fluences of $\rm 3 \times 10^{14}$, $\rm 6 \times 10^{14}$ and $\rm 1.2 \times 10^{15}\; /cm^{2}$ were studied. These fluences gave initial as-implanted peak hydrogen concentrations of $\rm 1 \times 10^{19}$, $\rm 2 \times 10^{19}$ and $\rm 4 \times 10^{19}\; /cm^{3}$. The peak concentration of these implants is expected to decrease during the SPE anneal as the H diffuses and the H concentration profile becomes broader. Fig.~\ref{ge_h_vel} shows the SPE velocity profiles for each of these three samples as well as the theoretical as-implanted H concentration profile. Also shown is the intrinsic SPE rate which is relatively constant except for depths less than 0.3~$\mu$m where retardation due to H indiffusion occurs. 

As can be seen, the interface velocity quickly drops to less than half that of the unimplanted sample when it encounters the hydrogen profile. This holds true for all three hydrogen fluences. It should be noted that after that point, the observed interface velocity cannot be directly correlated with the as-implanted H profile since it is expected that the H will exhibit diffusion broadening during SPE and that the interface will cause a redistribution of the H as was observed for the indiffused H (Fig.~\ref{ge_h_sims1}). However, some observations can be made based on the total hydrogen content in the sample if one assumes  that there is no large scale loss of H out through the surface of the sample or across the {\em c--a} interface during the initial stages of the annealing process. For example, the degree of retardation of the SPE rate between the $\rm 1 \times 10^{19}\; H/cm^{3}$ sample and the unimplanted case is large with respect to the difference between the $\rm 1 \times 10^{19}\; H/cm^{3}$ case and the other two higher implant fluences. This tends to indicate that there is a threshold or saturation level of the H concentration for which retardation occurs. This threshold behaviour is consistent with that observed for implanted hydrogen in $a$-Si as reported previously.\cite{ober,APL:roth,MRSSP:roth92}.

\begin{figure}
\rotatebox{0}{\includegraphics[height=12cm]{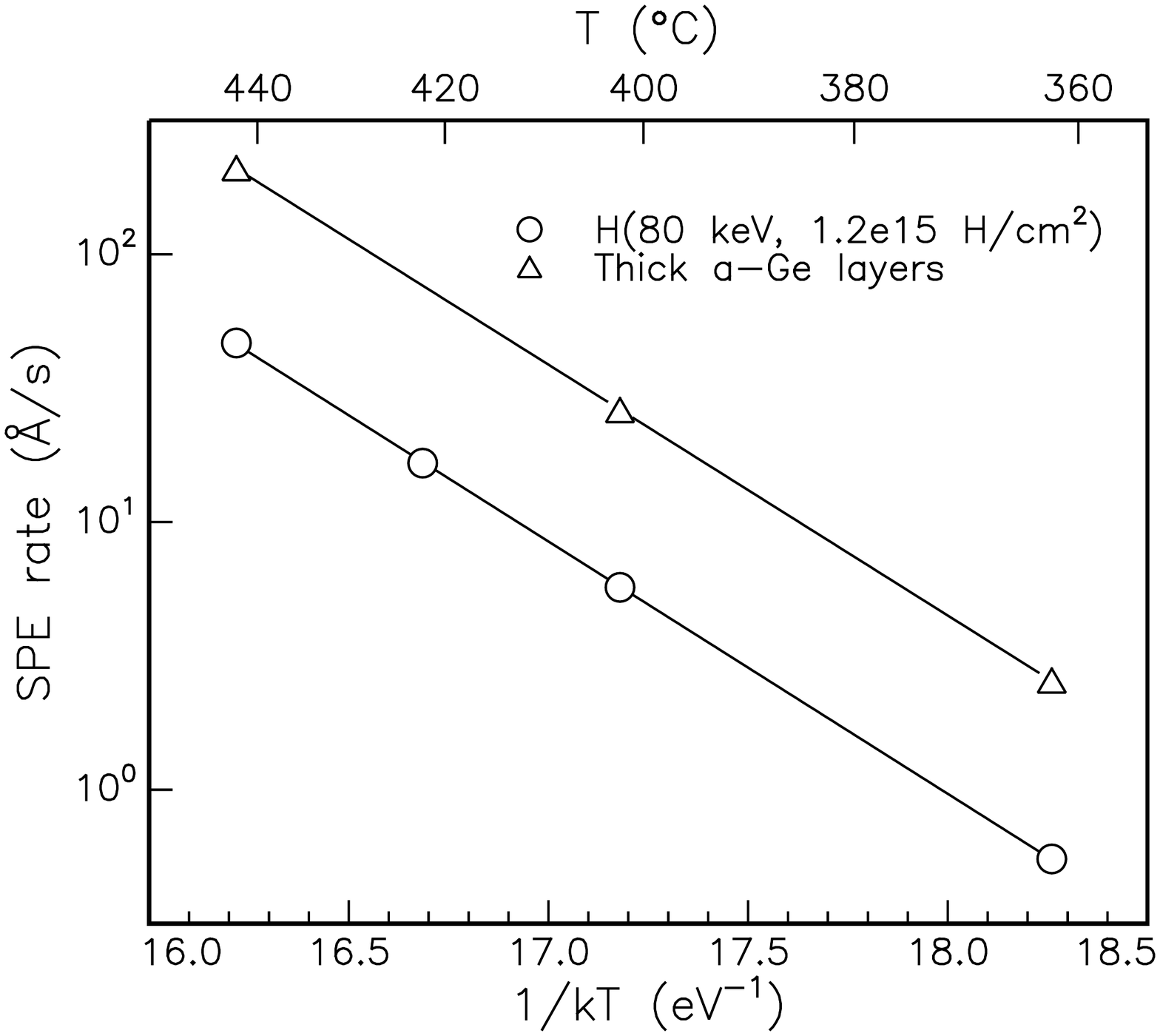}}
\caption[Arrhenius Dependence for~H Implanted $a$-Ge layers]{
The Arrhenius behavior of the SPE regrowth rate in 3.25~$\mu$m thick $a$-Ge layers ($\triangle$) determined in a H-free region and H implanted $a$-Ge ($\circ$). The H implant was to a fluence of $1.2\times 10^{15}\; cm^{2}$ with 80~keV H ions. The SPE velocity in these samples was taken as the mean velocity in the region from 0.2 to 0.5~$\mu$m. Arrhenius fits to the H-free and H implanted data yield 2.16~eV, $3.3\times 10^{7}\; m/s$ and 2.17~eV, $8.9\times 10^{6}\; m/s$, respectively.
}
\label{ge_h_arrh}
\end{figure}

In Fig.~\ref{ge_h_vel} the interface velocity is observed to be fairly constant in the region from 0.2~$\mu$m to 0.5~$\mu$m for all three hydrogen implanted cases. The mean velocity in this region for the sample implanted with $1.2 \times 10^{15}$~H/cm$^2$ is plotted at various temperatures in Fig.~\ref{ge_h_arrh} along with the H-free SPE data from the thick $a$-Ge layers described in the previous section. The activation energy and velocity prefactor found from fitting this data with an Arrhenius type equation for the H implanted sample are 2.17~eV and $8.9 \times 10^{6}$~m/s, respectively. Fits to the H-free data in the thick $a$-Ge sample yielded 2.16~eV and $3.3\times 10^{7}\; m/s$ respectively. 

It is evident from these fits that the addition of hydrogen does not alter the activation energy of the SPE process in $a$-Ge. However, the velocity prefactor is reduced by a factor of $\sim$3.7 from that of the H-free case. This result explains in part the wide variability of the velocity prefactor ($6.1\times 10^{6}$ to $7\times 10^{7}$~m/s) reported in the literature and also shows how these previous works were affected by the infiltration of H. 

The fact that the activation energy of the SPE process remains unchanged is consistent with that observed in the $a$-Si system as reported by Roth~{\em et al.}\cite{MRSSP:roth92} They further suggested that H passivates crystallization sites while not affecting the energy associated with a crystallization event. This also seems to be true of SPE in $a$-Ge layers.

In Si it has been found that the SPE rate decreases linearly with increasing interfacial hydrogen concentration up to approximately 3$\times 10^{19}$~H/cm$^3$, and that for concentrations beyond that there is little change in the SPE rate.\cite{review:roth} This threshold value has been correlated with the density of dangling bonds in ion-amorphized $a$-Si and this has been cited as evidence for the possible involvement of dangling bonds in the SPE process. While the results presented here are not sufficient to ascertain the exact concentration dependence for Ge, they do in fact provide bounding values on the hydrogen concentration threshold value at which point the SPE rate becomes essentially invariant for further increases in hydrogen content. From Fig.~\ref{ge_h_vel} it is evident that this threshold has been reached for the $4 \times 10^{19}$~H/cm$^3$ implant, which corresponds to an interfacial hydrogen concentration of $6 \times 10^{19}$~H/cm$^3$ as determined from SIMS (not shown).\cite{paul} The lower bound on this value is set by the interfacial concentration of 6$\times$10$^{18}$~H/cm$^{3}$ for indiffused H as observed in Fig.~\ref{ge_h_sims1}. In this case the SPE rate is still a strong function of the hydrogen concentration as displayed in Fig.~\ref{ge_h_vel}. Thus the hydrogen saturation concentration level in $a$-Ge lies somewhere between 6$\times$10$^{18}$~H/cm$^{3}$ and $6 \times 10^{19}$~H/cm$^3$. This value is comparable to the value  of $4 \times 10^{19}$~H/cm$^3$  reported by Roth {\em et al.}\ for hydrogen in silicon.\cite{MRSSP:roth92} However, in contrast to the Si system, the SPE rate at or near the H saturation level in Ge systems observed here is a factor of $\sim$3.7 times slower than the intrinsic value, as compared to the factor of $\sim$2 reported by Roth {\em et al.}\ for Si. The observation of the more efficient passivation of crystallization sites in $a$-Ge may lead to the development of better insight into the growth mechanisms in future. 

It is supposed that the lack of a significant H effect in $a$-Ge SPE data for intrinsic layers is due to the fact that Ge does not possess any significant oxide layer. Without the formation of a substantial oxide layer, there is not enough H available to penetrate into the $a$-Ge layer at a large enough concentration to cause SPE rate reduction.

\subsection{Dopant-enhanced SPE in $a$-Ge}
\label{dopant}

Multiple energy dopant implants were used to create three different constant concentration profiles of As or Al in $a$-Ge layers. The concentrations were $1\times 10^{19}$, $5\times 10^{19}$  and $10\times 10^{19}$~/cm$^{3}$ with a region of constant concentration covering a depth of 0.25--0.55~$\mu$m and 0.5--0.9~$\mu$m for As and Al, respectively. The SPE rates were determined for each dopant implanted sample within these depth ranges.

The accuracy of the concentration was determined by RBS measurements on As implanted Si samples that were prepared at the same time and under similar conditions as the As implanted Ge. The concentrations measured by RBS agreed to within 5\% of the expected concentrations. SIMS measurements were used to verify both the As and Al concentration profiles. The SIMS profiles compared well in shape and depth scale to expected profiles calculated with the Profile code.\cite{profile} The absolute Al concentration was confirmed with the implanter dosimetry. 

\begin{figure}
\begin{center}
\rotatebox{0}{\includegraphics[height=12cm]{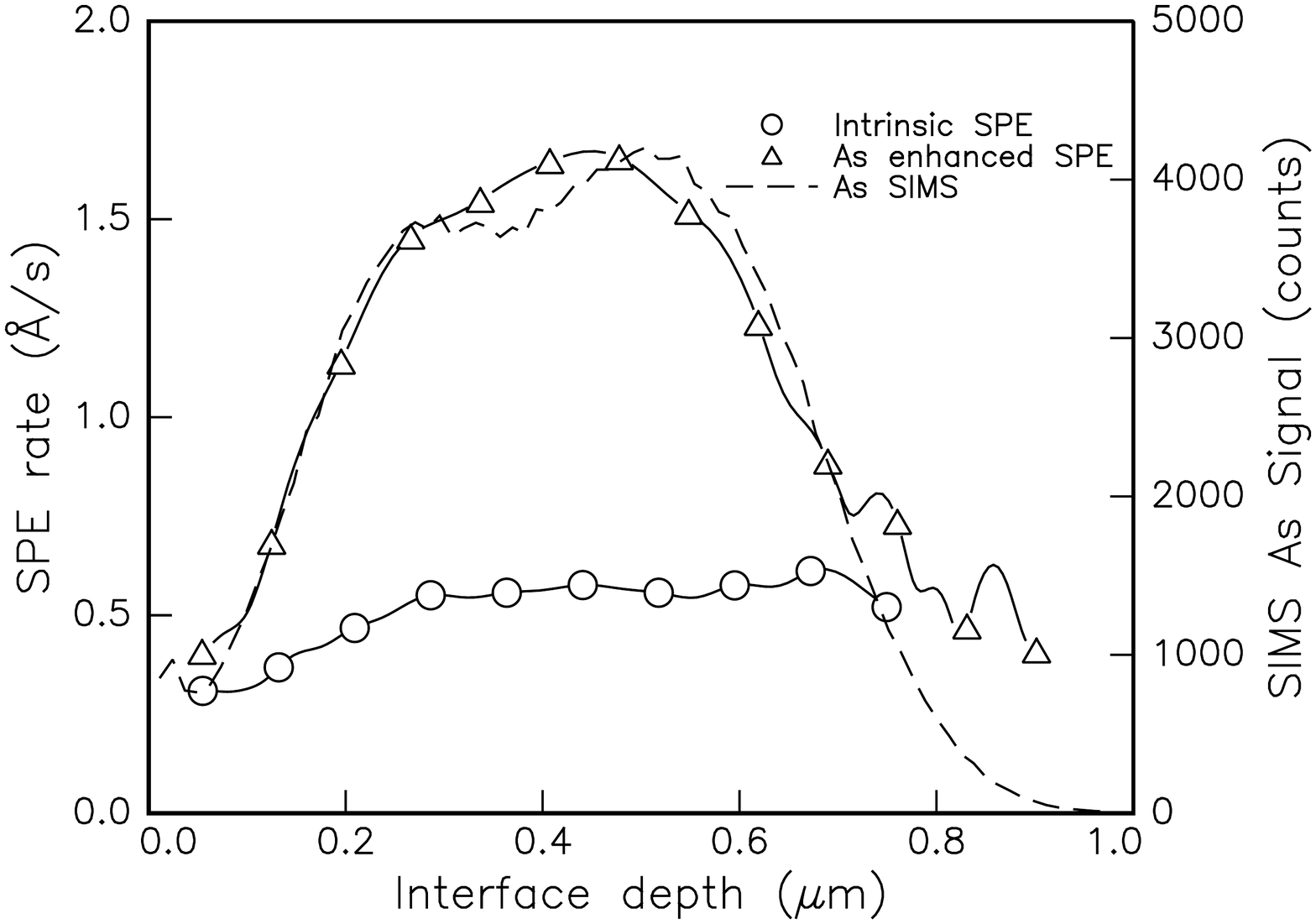}}
\end{center}
\caption{%
Comparison of the implanted As profile (dashed line) as determined from a SIMS measurement and the SPE rate enhancement ($\triangle$) due to the implanted As for crystallization at 340~$^{\circ}$C. The depth scale for the SIMS profile comes from a measurement of the sputter crater depth, whereas the depth scale assigned to the SPE profile relies on the index of refraction determined from the $a$-Ge. The SPE rate for an intrinsic $a$-Ge layer sample ($\circ$) is also shown for comparison.
}
\label{001}
\end{figure}

Figure~\ref{001} shows the As enhanced SPE rate for crystallization at 340~$^{\circ}$C compared to the implanted As concentration profile determined by SIMS. The SPE velocity and the concentration profiles agree well. Also shown is the intrinsic SPE rate. Hydrogen retardation can again be observed at depths less than 0.3~$\mu$m. 

\begin{figure}
\begin{center}
\rotatebox{0}{\includegraphics[height=8cm]{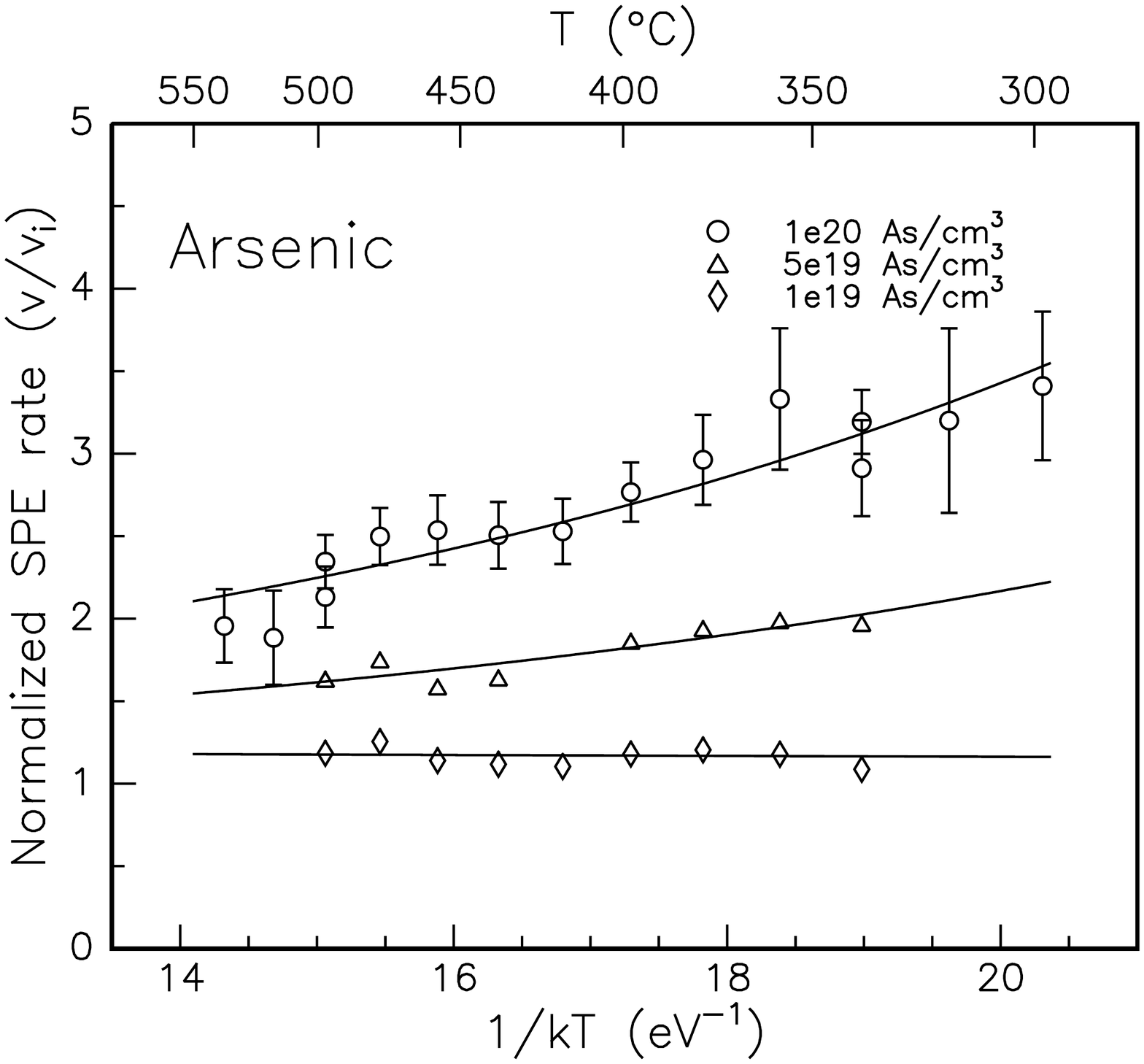}}
\rotatebox{0}{\includegraphics[height=8cm]{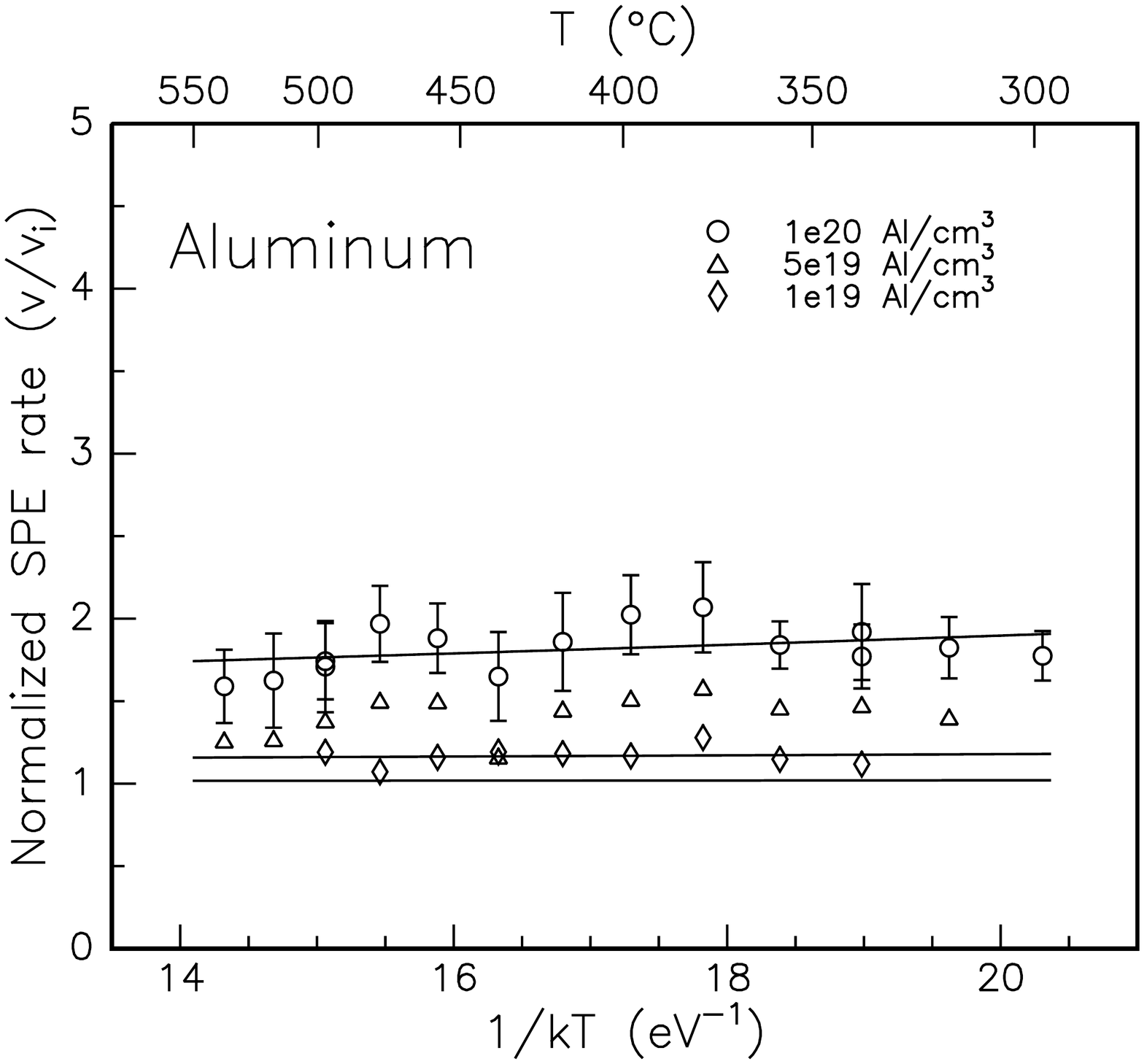}}
\end{center}
\caption{%
The SPE rates of As implanted (top panel) and Al implanted (lower panel) $a$-Ge are shown normalized against the SPE velocity of undoped $a$-Ge at each temperature. Three different concentrations are shown: $1\times 10^{19}$ ($\Diamond$), $5\times 10^{19}$ ($\triangle$) and $10\times 10^{19}$~/cm$^{3}$ ($\circ$). Solid lines are best fits to the data assuming a $v/v_{i}=1+N_{imp}/N_{i}$ dependence.
}
\label{002}
\end{figure}

Figure~\ref{002} shows the SPE rates normalized to the intrinsic rates at each temperature. The error bars shown in Fig.~\ref{002} were calculated by considering the reproducibility of the data and the RMS noise in the determined velocity curves. Error bars for the other two concentrations are of similar magnitude and have been omitted for clarity.

Such normalized velocity plots lend themselves well to comparison against SPE growth models which express their predictions in a similar format, such as the Fractional Ionization~(FI) Model of the Walser group,\cite{jeon1,MRSSP:walser} and the Generalized Fermi-Level Shifting~(GFLS) Model.\cite{lu,TOBE:gfls} For both of these models, the normalized SPE rate can be expressed in the general form,

\begin{equation}
\frac{v}{v_i}=1 + \frac {N_{impl}}{N_i}\  \  {\rm with} \  
\  N_i=N_o \, exp \, (- \Delta E/kT)
\label{wal_eqn}
\end{equation}

\noindent where $N_{impl}$ is the dopant concentration. The interpretation of $N_o$ and~$\Delta E$ is different for each of the FI and GFLS models, and it is this difference that may make one model more favorable than the other. In the FI model the rate limiting step to the SPE process is thought to be the capture of dangling bonds at the interface. The concentration of dangling bonds is determined by the band structure on the amorphous side of the interface where the Fermi level is pinned to midgap. The dangling bond concentration cannot then be modified by changes in the Fermi level directly. Instead, the dangling bond concentration is changed by ionization enhanced atomic mobility as per Bourgoin and Germain.\cite{PL:bourgoin} However, the assumption that the fractional ionization is independent of the doping concentration has no prior justification and does not take into account the law of mass action.\cite{lu} Therefore, compensation effects cannot easily be explained with this model. Although the FI model does not offer a clear explanation of the SPE process, Eq.~\ref{wal_eqn} does provide reasonable fits to the data.

The Walser group\cite{jeon4,park,walrus,jeon1} typically plotted $V/V_i$ against $N_{impl}/N_i$ when comparing data for Si to the predictions of Eq.~\ref{wal_eqn}. While this method of displaying the data is useful to verify that an extrapolation of the data goes through the expected point of~$(0,1)$ it obscures the temperature dependence of the SPE enhancement for a given doping level. The temperature dependence is easily seen in a normalized velocity plot as shown in Fig.~\ref{002} and is similar in form and magnitude to dopant enhanced SPE measurements performed in $a$-Si.\cite{NIMB:mccallum,TOBE:gfls}

\begin{table}
\begin{center}
\begin{tabular}{ccc}
$N_{impl}$ (10$^{20}$~/cm$^3$) &	$\Delta E$ (eV) & $N_o$ (10$^{20}$~/cm$^3$) \\  \hline
1		& 0.133$\pm$0.015 & 5.9$\pm$1.6 \\
0.5		& 0.14$\pm$0.03   & 7.6$\pm$4.7 \\
0.1		& 0.007$\pm$0.09  & 0.7$\pm$1.2 \\ \hline
\end{tabular}
\end{center}
\caption[SPE Enhancement Fluence Dependence for Arsenic in Ge]{
Fitting parameters to the experimental arsenic-enhanced SPE rates assuming a functional form as per Eq.~\ref{wal_eqn}. Error values are based on the estimated error from the fitting process. In the lowest fluence case, $N_{impl}$ is sufficiently small as to allow a wide range of parameters to match the data.}
\label{ge_as_norm_fit}
\end{table}

The best fits of Eq.~\ref{wal_eqn} to the data are shown as solid lines in Fig.~\ref{002}, with the associated~$\Delta E$ and $N_o$ values for the As data listed in Table~\ref{ge_as_norm_fit}. For the lowest fluence of only $1\times 10^{19}$~As/cm$^{3}$, within errors there is essentially no SPE rate enhancement evident, or in terms of Eq.~\ref{wal_eqn}, the $N_{impl}$ quantity is too small to have a significant effect. Hence a wide range of~$\Delta E$ and $N_o$ parameters will fit that data set. In contrast, the values of~$\Delta E$ and $N_o$ determined for the other two higher As concentrations are reasonably consistent with each other.

Suni~{\em et al.\/} observed SPE rate enhancements on the order of~1.5$\times$ for their measurements of $\sim1\times 10^{20}$~As/cm$^{3}$ in $a$-Ge at the two temperatures of~300$^{\circ}$C and 325$^{\circ}$C.\cite{suni2} The $1\times 10^{20}$~As/cm$^{3}$ case from this study, as shown in Fig.~\ref{002} exhibits at least a three times enhancement over the undoped SPE rate in that temperature range. One contributing factor to the difference in observed enhancements may be due to the fact that the region containing As for which the mean SPE rate was determined by Suni~{\em et al.}\ had less than the expected $1\times 10^{20}$~As/cm$^{3}$. No determination or verification of the actual concentration was reported. Our theoretical calculations of their implant given the implantation parameters puts the As concentration between $\sim8.5 \times 10^{19}$~As/cm$^{3}$ and $\sim9.7 \times 10^{19}$~As/cm$^{3}$ over the depth range 0.075 to 0.3~$\mu$m.

The difference between the SPE rate enhancement of Al and As is evident when the aluminum SPE data are plotted in normalized form, as in Fig.~\ref{002}. The normalized As data~(top panel of Fig.~\ref{002}) exhibited a reasonably clear temperature dependence for a given doping level, whereas the temperature dependence of the normalized Al SPE data is not as obvious. From the fits to the Al data it appears that the SPE rate enhancement for a fixed doping concentration also increases with decreasing temperature, but the trend is so slight that there is little value in reporting the parameters determined from fitting a function of the form of Eq.~\ref{wal_eqn}. Aluminum enhanced SPE in $a$-Si also shows only a weak temperature dependence.\cite{TOBE:gfls}

\begin{figure}
\begin{center}
\rotatebox{0}{\includegraphics[height=12cm]{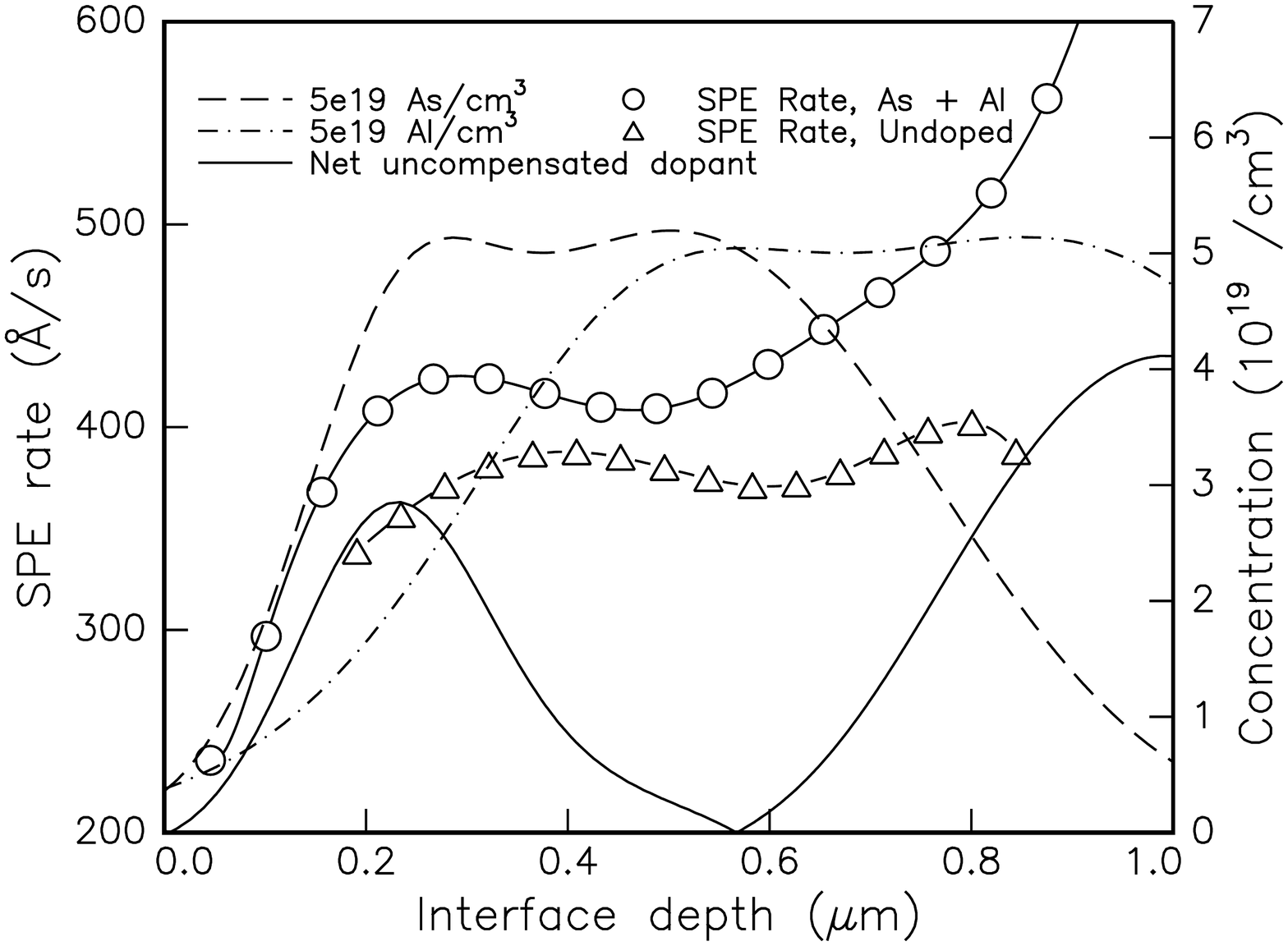}}
\end{center}
\caption{%
The SPE rate in compensation doped Ge ($\circ$), as determined from the interface location as a function of time. The two dopant profiles for As and Al combine to give a net dopant profile (solid line). The SPE rate in undoped Ge ($\triangle$) at the same temperature (460~$^{\circ}$C) is shown for comparison.
}
\label{004}
\end{figure}

A sample containing both As and Al concentrations at $\rm 5 \times 10^{19}\; /cm^{3}$ was also considered. The expected dopant profiles are shown in Fig.~\ref{004}. The expected net uncompensated dopant concentration is indicated by the solid line. The As and Al concentrations are equal at a depth of $\sim$0.56~$\mu$m. This results in an uncompensated dopant concentration of less than $5\times 10^{18}$~/cm$^{3}$ in the depth region from $\sim$0.4~$\mu$m to $\sim$0.6~$\mu$m. The SPE rates for these samples were investigated following the same procedure that was employed for the samples implanted with only As or Al.

For the depth range from $\sim$0.82~$\mu$m to $\sim$1.2~$\mu$m, the net uncompensated concentration is greater than 2.8$\times 10^{19}$~Al/cm$^{3}$ and the SPE rate is enhanced in this region compared to that of an undoped sample. As can be seen in Fig.~\ref{004}, for depths greater than $\sim$0.7~$\mu$m the SPE rate in the compensated sample increases as the uncompensated component increases. In the depth range 0.4--0.6~$\mu$m the SPE rate approaches the intrinsic rate and then diverges for depths less than 0.4~$\mu$m as the uncompensated component increases again.

For the purpose of correlating SPE rates to dopant compensation in $a$-Ge the interface velocity was taken as the mean velocity within the region between $\sim$0.44~$\mu$m to $\sim$0.64~$\mu$m. 
From Fig.~\ref{004} this corresponds to the range over which the net doping level will be less than 5$\times 10^{18}$~/cm$^{3}$ and it is evident that the SPE rate is essentially constant within this depth window. 

\begin{figure}
\begin{center}
\rotatebox{0}{\includegraphics[height=12cm]{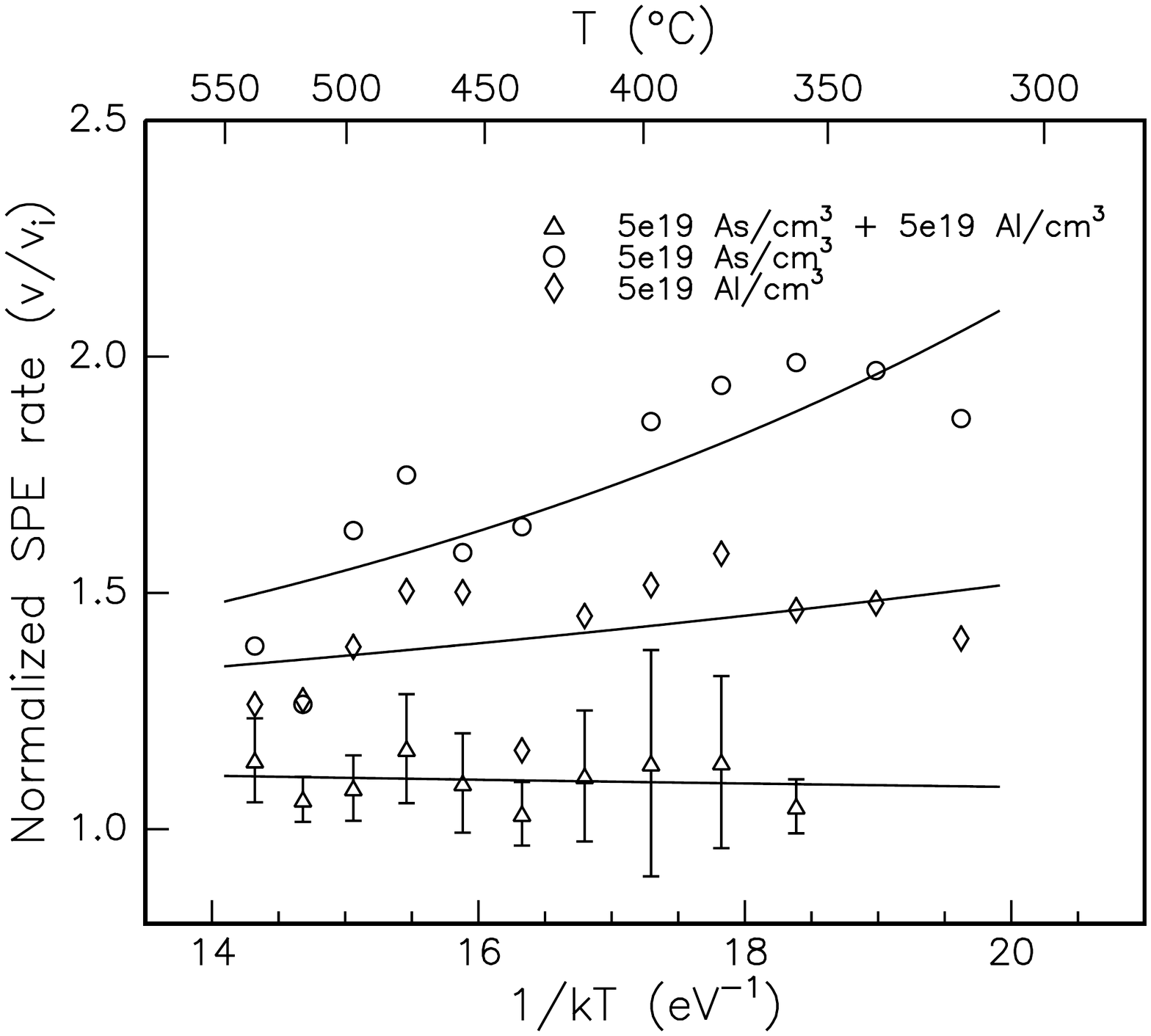}}
\end{center}
\caption{%
The SPE rates in compensation doped Ge ($\triangle$) as a function of temperature after being normalized to the undoped SPE rates. The normalized SPE rates associated with samples containing the same concentration of either As ($\circ$) or Al ($\Diamond$), but not both simultaneously, are shown for comparison.
}
\label{005}
\end{figure}

Once again, the significance of the difference between the compensated samples and the undoped samples is best observed on a normalized velocity plot such as that shown in Fig.~\ref{005}. Also shown is the As and Al enhanced SPE rates for concentrations of $5 \times 10^{19}$~/cm$^{3}$ from Fig.~\ref{002} for comparison. It can be seen that when the two dopants are combined in the same sample the SPE rate returns to a value comparable to that observed for the undoped material. This compensation behaviour is similar to that reported by Suni~{\em et al.}\ for overlapping single implants of B and As in $a$-Ge at peak concentrations on the order of~1$\times 10^{20}$~/cm$^{3}$.\cite{suni2} The slight offset of the normalized velocity in Fig.~\ref{005} for the compensation doped samples from~1.0 for intrinsic material to $\sim$1.1 may be due to the error in the concentrations for the two dopants in the compensation sample.

\section{Discussion}

\subsection{Intrinsic $a$-Ge and the Kinetic Model}
\label{kinetic}

The SPE process can be treated as a thermally excited transition from the amorphous to crystalline phase using transition state theory. Lu~{\it et~al.\/} have extended this theory by reconsidering the dangling bond model of Spaepen and Turnbull resulting in an extended kinetic theory of SPE.\cite{lu} This theory allows some comparisons to be made with the undoped Ge data presented in section~\ref{intrinsic}. Within the context of this model, the velocity prefactor is given by

\begin{equation}
V_\circ=2\sin(\theta)v_s n_r \exp\left(
\frac{\Delta S_f + \Delta S_m}{k}
\right)
\end{equation}

\noindent where $\theta$ is the misorientation from \{111\} (55$^\circ$ from the (100) surface), $v_s$ is the speed of sound, $n_r$ is the net number of jumps a dangling bond makes before it is annihilated, $\Delta S_f$ is the entropy of formation of a dangling bond pair, and $\Delta S_m$ is
the entropy of motion of the dangling bond at the interface.

By using the same bounds on the entropy terms as used by Lu~{\em et~al.\/}  it is possible to refine their estimate for the number of crystallization events per formation of a dangling bond pair, $N_r = 2\, r\, n_r$, where $r$ is the ratio of crystallization events to configurational coordinate steps.  The factor of two arises since $n_r$ is per dangling bond, whereas $N_r$ is per dangling bond pair. Substituting in the value of $V_\circ$ from this work, and using the same $r=3/9$ value\cite{saito84} and $v_s$ as was used by Lu~{\em et~al.\/} results in~$65\leq N_r \leq 2600$. In this case, the additional uncertainty associated with the velocity prefactor value is only $\sim$20\% instead of the previous amount of $\sim$50$\times$ estimated by Lu. The question of whether this is a reasonable number of crystallization events is still open to speculation bit at least now the experimental data place reasonably tight constraints on the possible range.

\subsection{Doped-Ge and the GFLS Model}
\label{gfls}

The generalized Fermi level shifting model links structural changes related to SPE at the interface to shifts in the Fermi level.\cite{TOBE:gfls} It has been applied to Si with some success. However,  application of the model to another material system, in this case Ge, would greatly add to confidence in the accuracy of the model in describing doping effects in the SPE growth mechanism. For the dopant concentrations used in this work, the Fermi levels were calculated numerically since Ge must be treated as a degenerate semiconductor for the temperatures and concentrations used in this work as shown in the appendix. Once the Fermi levels are known the normalized SPE velocity data for $n$-type material can be fitted using

\begin{equation} 
\label{fit}
\frac{v}{v_{i}} = \frac{1+g\exp\Bigl(\frac{E_{f}-E_{k}}{kT}\Bigr)}{1+g\exp\Bigl(\frac{E_{fi}-E_{k}}{kT}\Bigr)} 
\end{equation} 

\noindent where $E_{f}$ is the Fermi level and $E_{k}$ represents the energy level within the band gap of the defect responsible for the SPE process. The degeneracy factor, $g$, associated with $E_{k}$ is given by $g=Z(D^{-})/Z(D^{0})$ where $Z(D^{-})$ and $Z(D^{0})$ are the internal degeneracies of the $D^{-}$ and $D^{0}$ defect states, respectively.\cite{book:bourgoin} If a dangling bond defect is responsible for the SPE process then it is expected that $g=1/2$ if only the spin degeneracy needs to be considered. For the positive charge state of the dangling bond, $g=1$ as the degeneracy of the valence band also contributes a factor of two. The reduction of Eq.~\ref{fit} to Eq.~\ref{wal_eqn} requires a number of assumptions and approximations as performed by Lu~{\em et al.}\cite{lu} For example, assuming that the Fermi level in intrinsic material is at midgap.  

The normalized velocity data for the As implanted samples shown in Fig.~\ref{002} exhibit a larger temperature variation than the Al implanted samples. This enables more accurate fits to be performed and therefore, only fits to these data will be considered for the purposes of the following discussion. 

By allowing both $g$ and $E_{k}$ to vary in Eq.~\ref{fit}, fits to the normalized As SPE rates were of equal quality to the fits for the fractional ionization model (Eq.~\ref{wal_eqn}). As before, a large range of fitting parameters could be used on the $\rm 1\times 10^{19}\; /cm^{3}$ data due to the fact that the velocity ratio is essentially invariant. The values of $E_{k}$, referenced to the conduction band edge, and $g$ obtained from fitting the $\rm 5\times 10^{19}\; /cm^{3}$ and $\rm 10\times 10^{19}\; /cm^{3}$ data sets are $(E_{c}-E_{k})=(0.07\pm0.03)$~eV and $g=0.4\pm0.2$ and $(E_{c}-E_{k})=(0.06\pm0.01)$~eV and $g=0.6\pm0.2$, respectively. The error values quoted here are from the fits only. This results in a SPE defect level that is about 0.06~eV below the conduction band edge and a degeneracy value of about a half. If the degeneracy factor is fixed to a value of a half and only $E_{k}$ is allowed to vary then one obtains $(E_{c}-E_{k})=0.053$~eV and $(E_{c}-E_{k})=0.072$~eV for the $\rm 5\times 10^{19}\; /cm^{3}$ and $\rm 10\times 10^{19}\; /cm^{3}$ data, respectively. The quality of the fits remains fairly reasonable in this case. 

The degeneracy value of $g=1/2$ is consistent with the negative charge state of a dangling bond type defect. This value also agrees with similar SPE studies in As implanted Si where a value of a half was obtained.\cite{TOBE:gfls} The defect energy level is relatively close to the conduction band edge. As a reference point for this energy level, other defects in $c$-Ge can be found at  $(E_{c}-0.04)$~eV tentatively assigned to the transition level of a self-interstitial,\cite{prl:haesslein} $(E_{c}-0.39)$~eV for a negatively charged divacancy and $(E_{c}-0.54)$~eV for a negatively charged vacancy.\cite{Coutinho:2006aa} The defect level found here by the GFLS model is entirely consistent with the band-gap positions of some known defects. However, to the best of our knowledge the band-gap state for a dangling bond in $c$-Ge is not in the literature. 

In $a$-Si layers, fits to As enhanced normalized SPE rates yielded $(E_{c}-0.16)$~eV which also compares well to the energy levels of typical charged defects in $c$-Si.\cite{TOBE:gfls} The consistency of the GFLS model for both the Si and Ge systems is thus demonstrated. 

A shortcoming to the GFLS model is that the design of an independent experiment to identify the SPE defect is difficult. It is interesting to note however, that some early molecular dynamics (MD) simulations also attributed the SPE mechanism to the motion of a dangling bond type defect.\cite{saito81,saito84} These simulations attempted to describe the structure and rearrangement of atoms at the $c$-$a$ interface during SPE on a microscopic scale using a simple harmonic potential. More recent MD simulations by Bernstein {\em et al.} have shown that the SPE may occur through a number of both simple and complex mechanisms.\cite{prb:bernstein98,prb:bernstein00} By using empirical potential simulations they have found that one simple mechanism involves the rotation of two atoms aided by coordination defects which are locally created and annihilated during crystallization. An example of a more complex mechanism involves the migration to the interface of a five-fold coordinated defect which aides the incorporation of two atoms into the crystal matrix. Each process taking part in SPE may have a different activation energy. Motooka {\em et al.} has also identified two different activation energies for low and high temperatures via MD simulations.\cite{PhysRevB.61.8537} Now, if these MD simulations accurately model the SPE process, doubt is cast on the generally accepted idea that SPE occurs through a single, thermally-activated process. 

To date only limited number of studies have considered dopants in MD simulations of the SPE process. One such study described the segregation and precipitation of B during SPE in highly doped Si.\cite{prb:mattoni} This is shown to result in the retardation of the SPE rate and is in agreement with experiment.\cite{review:roth} However, dopant-enhanced SPE is not considered. Indeed, all MD simulations are performed near the melting point of amorphous silicon in order to ensure reasonable simulation times. There are no MD simulations that we know of that have been performed in the temperature range considered in the present work, or previous work with Si,\cite{TOBE:gfls} where the effect of the dopants on the SPE rate becomes apparent (as can be seen at the lower temperatures in Fig.~\ref{002}). If such MD simulations become possible, dopant-enhanced SPE may be understood to a greater extent on the microscopic level and could then be used to assess the applicability of the GFLS model.

\section{Conclusions}
The H-free SPE regrowth rates of $a$-Ge layers formed by self-ion implantation of various $<$100$>$ Ge substrates obtained from different suppliers were measured at temperatures in the range 300--540{$\;^{\circ}$C}. From these measurements, an activation energy of (2.15$\pm$0.04)~eV with a velocity prefactor of~(2.6$\pm$0.5)$\rm \times10^{7}\; m/s$ was determined. No significant variation was observed between samples with different background doping levels. The amorphisation conditions were also varied, and again the SPE rate showed no observable change. 

The SPE rate was observed to slow down significantly when the {\em c--a} interface was within 0.3~$\mu$m of the surface. This was shown to be due to the infiltration of H into the substrate through the surface. In light of this, all prior measurements of SPE in Ge that include data from within this surface region are most likely not an accurate reflection of the bulk Ge SPE behavior. The infiltrating H does not penetrate as deeply into the $a$-Ge layers as in $a$-Si. It was reasoned that this is due to the inability of $a$-Ge to form a stable native oxide. Through H implantation it was shown that H has a greater effect on the SPE rate in $a$-Ge than it does in the $a$-Si system with the SPE rate being reduced by over 70\%. This observation may lead to the development of better insight into the SPE growth mechanisms in future. 

Measurement of the SPE rate in thick $a$-Ge layers has shown that it remains constant for {\em c--a} interface depths between 0.3~$\mu$m and 3.2~$\mu$m. The explosive crystallization effect becomes significant for $a$-Ge layers of this thickness, and it is suggested that SPE measurements involving thicker $a$-Ge layers would be impractical for device fabrication due to this problem.

To test the validity of SPE models such as the kinetic model and the GFLS model it is extremely important to broaden the range of material systems to which they are applied. The similarity between Ge and Si makes it one of the most obvious candidates. But, until now the SPE data for Ge was not sufficiently accurate (due largely to not accounting for H infiltration effects) for meaningful results to be obtained.  In applying the kinetic model, our improved data leads to an enormous reduction of the uncertainty, from a factor of 50 to 20\%, greatly increasing confidence in the applicability of the model.

The presence of As and Al to concentrations greater than $\rm 1\times 10^{19}\; /cm^{3}$ resulted in enhanced SPE rates similar to those found for Si. A compensation effect was also observed for $a$-Ge regions containing equal concentrations of As and Al where the SPE rate returned to a value close to the intrinsic value.

Both the fractional ionization model and the GFLS model yielded excellent fits to the dopant enhanced SPE data. Theoretical calculations demonstrated that the material would be degenerate for the dopant concentrations and temperatures used in this work. Therefore, non-degenerate simplifications cannot be used. 

The degeneracy and energy level of the SPE defect determined from the As enhanced SPE data via the GFLS model was $(E_{c}-E_{k})=0.07\pm0.01$~eV and $g=0.5\pm0.2$. These results are remarkably consistent with previous studies performed in $a$-Si and are also consistent with the possibility of a negatively charged dangling bond being involved in the SPE process. By showing that application of the model to a material system other than Si also yields plausible values for the defect states energy and degeneracy of the hypothesized SPE defect adds greatly to confidence in the predictive power of the model. 

\begin{acknowledgments}
The Department of Electronic Materials Engineering at the Australian National University is acknowledged for their support by providing access to SIMS and ion implanting facilities. This work was supported by a grant from the Australian Research Council.
\end{acknowledgments}


\appendix*
\section{}

\subsection{Fermi Levels in Germanium}

An integral part of applying the generalized Fermi level shifting (GFLS) model to our data is identifying a reliable parameter set and calculating the required Fermi levels. The Fermi level in both intrinsic and doped Ge needs to be determined in order to utilize  Eq.~\ref{fit} and fit the dopant enhanced SPE data. The carrier concentrations in the conduction and valence bands are respectively given by\cite{book:sze}

\begin{subequations}
\label{xi}
\begin{equation}
\label{electronconc}
n_{e}=\frac{2N_{c}}{\sqrt{\pi}}\mathcal{F}_{1/2}\Bigl(\eta\Bigr)  
\end{equation}
and
\begin{equation}
\label{holeconc}
n_{h}=\frac{2N_{v}}{\sqrt{\pi}}\mathcal{F}_{1/2}\Bigl(\eta\Bigr)
\end{equation}
\end{subequations}

\noindent where $N_{c}$ and $N_{v}$ are the effective DOS in the conduction and valence bands and $\eta = (E_f - E_c)/kT$ for $n_e$ and $\eta = (E_v - E_f)/kT$ for $n_h$. $E_{c}$ and $E_{v}$ are the energy levels of the conduction and valence band edges. The effective DOS in turn depends on the electron and the hole effective masses which are given by 0.56 and 0.35, respectively.\cite{kittel:tp} Effective mass values for Si have been found to have a temperature dependence.\cite{JAP:green} However, to the best of our knowledge, corresponding Ge data is not available at this time.

The Fermi-Dirac integral $F_{1/2}( \eta )$ in Eq.\ref{xi} can be approximated by its limiting form, $(\sqrt \pi /2) \, \exp( \eta )$, which is a good approximation for $\eta \leq -2$, but diverges rapidly from $F_{1/2}( \eta )$ for $\eta \geq -1$.\cite{kittel:tp} This usually occurs when the Fermi level comes to within $2kT$ of either of the band edges. For an intrinsic semiconductor the two carrier concentrations are equal. These equations can then be solved for the Fermi level, $E_{fi}$ using the limiting form to the Fermi-Dirac integral giving\cite{book:sze}

\begin{equation}
E_{fi}=\frac{E_{g}}{2}+\frac{k T}{2} \ln\Bigl(\frac{N_{v}}{N_{c}}\Bigr)
\end{equation}

\noindent where $E_{g}$ is the band gap energy referenced to the valence band edge and $N_{v}$ and $N_{c}$ are the effective density of states of the valence and conduction bands, respectively. The band gap energy has a temperature dependence which is often described by the semi-empirical formula given by Varshni,\cite{physica:varshni}

\begin{equation}
\label{varshni}
E_{g}=E_{o}-\frac{\alpha T^{2}}{T+\beta}
\end{equation}

\noindent where $E_{o}$ is the energy gap at $T=0$~K. For $c$-Ge, $E_{o}=0.7437$~eV,  $\alpha =4.774 \times 10^{-4}$~$\rm eV.K^{-1}$ and $\beta =235$~K.\cite{sze} These parameters were determined by fitting data only up to 177$^{\circ}$C. However, Thurmond notes that they accurately predict the experimental values at temperatures just below the melting point and thus they are expected to be valid over the entire temperature range.\cite{JES:thurmond} 

For a lightly doped $n$-type semiconductor, if the donor concentration, $N_{d}$, is large compared to the intrinsic carrier concentration, $n_{i}$, then it is a reasonable approximation to let $n_{e}$ equal the ionized donor concentration. Then the Fermi level can once again be determined with the limiting form of the Fermi-Dirac equation.

\begin{figure}
\begin{center}
\rotatebox{0}{\includegraphics[height=12cm]{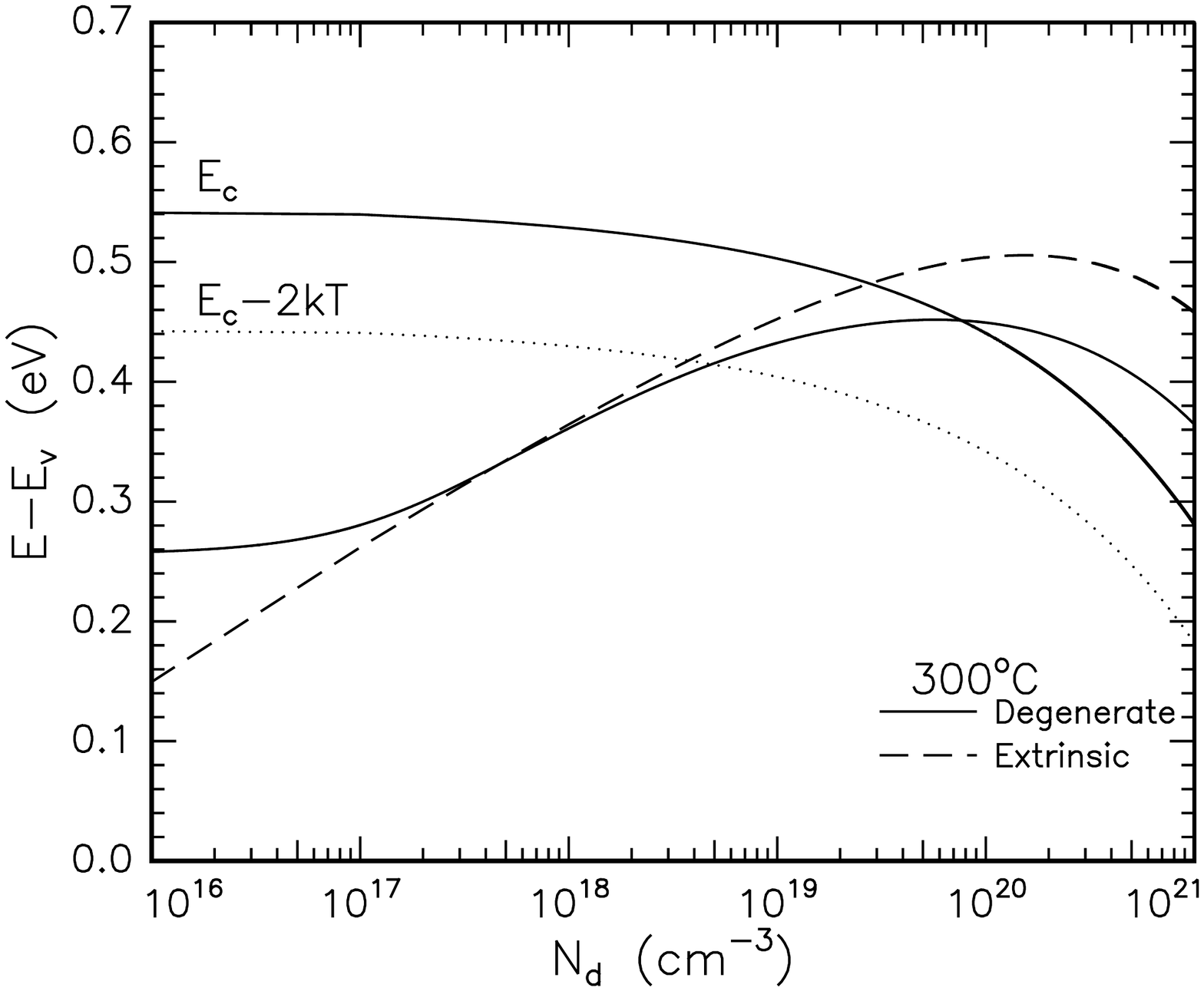}}
\end{center}
\caption{%
The Fermi level as a function of As concentration calculated by solving the electrical neutrality condition for a degenerate semiconductor. The dashed line represents the Fermi level calculated using nondegenerate semiconductor statistics. The dotted line represents a $2kT$ window beyond which a degenerate approach must be taken.
}
\label{bgn}
\end{figure}

Figure~\ref{bgn} shows the Fermi levels calculated using both the non-degenerate (dashed line) and degenerate (solid line) approaches as a function of the As dopant concentration at a temperature of 300$^{o}$C. This is the lowest temperature used in these experiments and, for a constant concentration, the Fermi level will be closest to the band edge at this temperature. The method used to calculate the degenerate Fermi level is described elsewhere.\cite{TOBE:gfls} 

Both Fermi levels agree within a dopant concentration range of about $\rm 1 \times 10^{17}-4 \times 10^{18}\; cm^{-3}$. In the lower concentration range ($<\rm 1 \times 10^{17} cm^{-3}$) the approximation that $n_{e}\simeq N_{d}$ is no longer appropriate as
carriers generated thermally will dominate the electrical properties of the semiconductor. In the high concentration regime ($>\rm 5 \times 10^{18}\; cm^{-3}$) the Fermi-Dirac limiting form cannot be used. Above this dopant concentration the Fermi level crosses over the 2$kT$ window and into the degenerate regime. In this concentration regime, band gap narrowing due to the dopant concentration also becomes apparent especially for As concentrations above $\sim\rm 1 \times 10^{20}\; cm^{-3}$.\cite{SSE:jain} 

A similar treatment of Ge doped with Al shows that for Al concentrations above $\sim 3 \times 10^{18}$~$\rm /cm^{3}$, the Fermi level can be expected to lie within $2kT$ of the valance band edge at 300~$^{\circ}$C. Hence a degenerate semiconductor treatment is required for all Fermi level calculations performed in this work.

\newpage

\newpage

\end{document}